\documentclass[pra,aps,twocolumn,floatfix,showpacs]{revtex4}
\usepackage{graphicx,graphics,psfrag,amsmath,calc}
\usepackage{epsfig}
\usepackage{color}
\begin{document}
\title{
Artificial spin-orbit coupling in ultra-cold Fermi superfluids
}

\author{Kangjun Seo, Li Han and C. A. R. S{\'a} de Melo}
\affiliation{School of Physics, Georgia Institute of Technology,
Atlanta, Georgia 30332, USA}
\date{\today}

\begin{abstract}
The control and understanding of interactions in many 
particle systems has been a major challenge in contemporary 
science, from atomic to condensed matter and astrophysics.
One of the most intriguing types of interactions is the 
so-called spin-orbit coupling - the coupling 
between the spin (rotation) of a particle and its momentum 
(orbital motion), which is omnipresent both in the macroscopic
and microscopic world. In astrophysics, the spin-orbit coupling 
is responsible for the synchronization of the rotation (spinning)
of the Moon and its orbit around Earth, such that we can only see one
face of our natural satellite. In atomic physics, the spin-orbit
coupling of electrons orbiting around the nucleus gives rise to
the atom's fine structure (small shifts in its energy levels). 
In condensed matter physics, spin-orbit effects are responsible
for exotic electronic phenomena in semiconductors 
(topological insulators) and in superconductors without 
inversion symmetry. Although spin-orbit coupling 
is ubiquitous in nature, it was not possible to control it
in any area of physics, until it was demonstrated 
in a breakthrough experiment~\cite{spielman-2011} that 
the spin of an atom could
be coupled to its center-of-mass motion by dressing two atomic
spin states with a pair of laser beams. This unprecedented
engineered spin-orbit coupling was produced in ultra-cold bosonic
atoms, but can also be created for ultra-cold 
fermionic atoms~\cite{spielman-2011, sinova-2009, chapman-sademelo-2011}. 
In anticipation of experiments, we develop a theory for interacting 
fermions in the presence of
spin-orbit coupling and Zeeman fields, and show
that many new superfluids phases, which are topological in
nature, emerge. Depending on values of spin-orbit coupling, Zeeman fields,
and interactions, initially gapped $s$-wave superfluids 
acquire $p$-wave, $d$-wave, $f$-wave and higher angular momentum components, 
which produce zeros in the excitation spectrum, rendering 
the superfluid gapless.
Several multi-critical points, which separate topological superfluid phases 
from normal or non-uniform, are accessible depending on spin-orbit
coupling, Zeeman fields or interactions, setting the stage for the study
of tunable topological superfluids.

\end{abstract}
\maketitle

%
%

The effects of spin-orbit coupling in few body systems like the Earth-Moon
complex in astrophysics or the electron spin and its orbital motion around 
the nucleus in isolated atoms of atomic physics are reasonably well understood
due to the simplificity of these systems.
However, in the setting of many identical particles, spin-orbit effects 
have revealed quite interesting surprises recently 
running from topological insulators in semiconductors~\cite{kane-2005} 
to exotic superconductivity~\cite{gorkov-2001} and
non-equillibrium effects~\cite{galitski-2007} depending on the precise form 
of the spin-orbit coupling. In atomic physics 
the coupling arises from the interaction of the magnetic moment 
of the electron and a magnetic field, present in the frame of electron, due 
to the electric field of the nucleus. Similarly in condensed matter physics, 
the coupling arises from the magnetic moment ${\bf m}$ 
of electrons, which move in the background of ions. 
In the electron's reference frame,
these ions are responsible for a magnetic field ${\bf B}$, 
which depends on the electron's momentum ${\bf k}$ and couple
to electron's spin. 
The resulting spin-orbit coupling has the form 
$ 
{H}_{SO}  
= 
-{\bf m} \cdot {\bf B}
=
-
\sum_j 
h_j ({\bf k}) \sigma_j
,
$
where $\sigma_j$ represents the  
Pauli matrices and $h_j ({\bf k})$
describes the $j$-th component $(j = x, y, z)$ of  
the effective magnetic field vector ${\bf h}$. 
For some materials ${\bf h}$ can take
the Dresselhaus~\cite{dresselhaus-1955} form 
$
{\bf h}_D ({\bf k})
= 
v_D 
(
k_y {\hat {\bf x}} 
+
k_x {\hat {\bf y}}
), 
$
the Rashba~\cite{rashba-1984} form 
$
{\bf h}_R ({\bf k})
= 
v_R
(
- k_y {\hat {\bf x}} 
+
k_x {\hat {\bf y}}
), 
$
or more generally  
a linear combination of the two
$
{\bf h}_{\perp} ({\bf k})
= 
{\bf h}_{D} ({\bf k}) + {\bf h}_R ({\bf k}).
$
In all these situations the type of spin-orbit coupling 
can not be changed arbitrarily and the magnitude 
can not be tuned from weak to strong, making the 
experimental control of spin-orbit effects 
very difficult.

Recently, however, it has been demonstrated experimentally 
that spin-orbit coupling can be engineered in a ultra-cold 
gas of bosonic atoms in their Bose-Einstein 
condensate phase~\cite{spielman-2011}, 
when a pair of Raman lasers creates a coupling between 
two internal spin states of the atoms and 
its center-of-mass motion (momentum).
Thus far, the type of spin-orbit field that has been created in the 
laboratory~\cite{spielman-2011} has the 
equal-Rashba-Dresselhaus (ERD) form 
${\bf h}_{\perp} ({\bf k}) = {\bf h}_{ERD} ({\bf k}) = v {k_x} {\hat{\bf y}}$, 
where $v_R = v_D = v/2$. Other forms of spin-orbit fields 
require additional lasers and create further experimental 
difficulties~\cite{dalibard-2010}.
In ultra-cold bosons the momentum-dependent 
ERD coupling has been created in conjunction
with uniform Zeeman terms, which are independent of momentum, 
along the z axis (controlled by the Raman coupling $\Omega_R$),
and along the y-axis (controled by the detuning $\delta$). 
The simultaneous presence of $h_z$, $h_y$ and 
${\bf h}_{ERD} ({\bf k})$ leads
to the Zeeman-spin-orbit (ZSO) Hamiltonian
$$
H_{ZSO} ({\bf k})
= 
- h_z \sigma_z - h_y \sigma_y - h_{ERD} ({\bf k}) \sigma_y
$$
for an atom with center-of-mass momentum ${\bf k}$ and 
spin basis $\vert\uparrow \rangle$, 
$\vert\downarrow \rangle$.  
The fields 
$h_z = - \Omega_R/2$,  $h_y = -\delta/2$ 
and ${\bf h}_{ERD} = vk_x {\hat y}$ can be controlled independently,
and thus can be used as tunable parameters to explore the available 
phase space and to investigate phase transitions, 
as achieved in the  experiment
involving a bosonic isotope of Rubidium ($^{87}$Rb). Although
current experiments have focused on Bose atoms, there is 
no fundamental reason that impeeds the realization of a similar
set up for Fermi atoms~\cite{spielman-2011, sinova-2009, chapman-sademelo-2011}
designed to study fermionic superfluidity~\cite{chapman-sademelo-2011}. 
Considering possible experiments with 
fermionic atoms such as $^6$Li, $^{40}$K, 
we discuss in this letter phase diagrams, topological phase
transitions, spectroscopic and thermodynamic properties
at zero and finite temperatures 
during the evolution from BCS to BEC superfluidity 
in the presence of controllable Zeeman and 
spin-orbit fields in three dimensions.

To investigate artificial spin-orbit and Zeeman fields 
in ultra-cold Fermi superfluids, we start from the Hamiltonian
density
\begin{equation}
\label{eqn:hamiltonian}
{\cal H} ({\bf r})
=
{\cal H}_0 ({\bf r})
+
{\cal H}_I ({\bf r}),
\end{equation}
where the single-particle term is simply
\begin{equation}
\label{eqn:hamiltonian-single-particle} 
{\cal H}_0 ({\bf r}) 
=
\sum_{\alpha \beta} 
\psi^{\dagger}_{\alpha} ({\bf r}) 
\left[ 
{\hat K}_{\alpha} \delta_{\alpha \beta} 
- 
\sum_j {\hat h}_j ({\bf r})\sigma_{j,\alpha\beta} 
\right] 
\psi_{\beta} ({\bf r}).
\end{equation}
Here, $ {\hat K}_{\alpha} = - \nabla^2/(2 m) - \mu_{\alpha}
$ is the kinetic energy in reference to the chemical potential
$\mu_{\alpha}$, ${\hat h}_j ({\bf r})$ is the 
combined effective field including Zeeman and
spin-orbit components along the $j$-direction 
$(j = x, y, z)$, and $\psi^\dagger_{\alpha} ({\bf r})$
are creation operators for fermions 
with spin $\alpha$ at position ${\bf r}$.
Notice that we allow the chemical potential $\mu_\uparrow$
to be different from $\mu_\downarrow$, such that the number
of fermions $N_\uparrow$ with spin $\uparrow$ may be different
from the number of fermions with spin $\downarrow$. 
The interaction term is
\begin{equation}
\label{eqn:hamiltonian-interaction}
{\cal H}_I ({\bf r})
=
-g
\psi^{\dagger}_{\uparrow} ({\bf r})
\psi^{\dagger}_{\downarrow} ({\bf r})
\psi_{\downarrow} ({\bf r})
\psi_{\uparrow} ({\bf r}),
\end{equation}
where $g$ represents a contact interaction that can be 
expressed in terms of the scattering length
via the Lippman-Schwinger relation
$
V/g 
= 
-V m/(4\pi a_s) 
+ 
\sum_{\bf k} 1/(2\epsilon_{\bf k}).
$
The introduction of the average pairing field 
$
\Delta ({\bf r})
\equiv
g
\langle
\psi_{\downarrow} ({\bf r})
\psi_{\uparrow} ({\bf r})
\rangle
\approx \Delta_0
$
and its spatio-temporal fluctuation $\eta({\bf r}, \tau)$ 
produce a complete theory for superfluidity in this system.

From now on, we focus on the 
experimental case where a) the Raman detuning is zero 
$(\delta = 0)$ indicating that there is 
no component of the Zeeman field along the $y$ direction;
b) the Raman coupling $\Omega_R$ is non-zero meaning that a Zeeman
component along the $z$ direction exists, that is, 
$h_z = - \Omega_R/2$; and 
c) the spin-orbit field has components 
$h_y ({\bf k})$ and $h_x ({\bf k})$ along the
$y$ and $x$ directions. To start our discussion, we neglect 
fluctuations, and transform $H_0 ({\bf r})$ into momentum
space as $H_0 ({\bf k})$. Using the basis
$
\psi_{\uparrow}^\dagger ({\bf k}) \vert 0\rangle
\equiv \vert {\bf k} \uparrow \rangle,
$ 
$
\psi_{\downarrow}^\dagger ({\bf k}) 
\vert 0 \rangle
\equiv \vert {\bf k} \downarrow \rangle,
$ 
where $\vert 0\rangle$ is the vacuum state, the Fourier-transformed
Hamiltonian $H_0 ({\bf k})$ becomes the matrix 
$$
{\bf H}_{0} ({\bf k}) 
=
K_+ ({\bf k}) {\bf 1}
+
K_- \sigma_z
- h_z \sigma_z
- h_y ({\bf k}) \sigma_y
- h_x ({\bf k}) \sigma_x,
$$
Such matrix can be diagonalized 
in the helicity basis 
$
\Phi_{\Uparrow}^\dagger ({\bf k}) \vert 0 \rangle
\equiv \vert {\bf k} \Uparrow \rangle,
$
$
\Phi_{\Downarrow}^\dagger ({\bf k}) \vert 0 \rangle
\equiv \vert {\bf k} \Downarrow \rangle,
$ 
where the spins $\Uparrow$ and $\Downarrow$ are aligned or
antialigned with respect to the effective magnetic field 
$
{\bf h}_{\rm eff} ({\bf k}) 
= 
{\bf h}_{\parallel} ({\bf k}) 
+
{\bf h}_{\perp} ({\bf k}).
$
Here,
$
K_+ ({\bf k}) 
= 
(K_\uparrow + K_\downarrow)/2 
= 
\epsilon_{\bf k} 
- 
\mu_+,$
is a measure of the average kinetic energy
$
\epsilon_{\bf k} = k^2/2m
$
in relation to the average chemical
potential 
$
\mu_+ 
= 
(\mu_\uparrow + \mu_\downarrow)/2.
$
While
$
{\bf h}_{\perp} ({\bf k})
=
h_x ({\bf k}) \hat{\bf x}
+
h_y ({\bf k}) \hat{\bf y}
$
is the spin-orbit field
and
$
{\bf h}_\parallel ({\bf k}) 
=
(h_z - K_-) \hat{\bf z}
$
is the effective Zeeman field, with 
$
K_- 
= 
(K_\uparrow - K_\downarrow)/2  
= 
- \mu_- 
$
where 
$
\mu_- 
= 
(
\mu_\uparrow 
- 
\mu_\downarrow
)
/
2
$ 
is the internal Zeeman field due to initial 
population imbalance, and $h_z$ is the external 
Zeeman field. When there is no population imbalance
the internal Zeeman field is $\mu_- = 0$, and we have 
only $h_z$. In general, the 
eigenvalues of the Hamiltonian matrix ${\bf H}_0 ({\bf k})$ are 
$
\xi_{\Uparrow} ({\bf k})  
= 
K_+({\bf k}) - \vert {\bf h}_{\rm eff} ({\bf k})\vert  
$
and
$
\xi_{\Downarrow} ({\bf k})  
= 
K_+({\bf k}) + \vert {\bf h}_{\rm eff} ({\bf k})\vert,  
$
where
$
\vert {\bf h}_{\rm eff}({\bf k})\vert  
=
\sqrt{
(\mu_- + h_z)^2 
+ 
\vert h_\perp ({\bf k}) \vert^2
}
$
is the magnitude of the effective magnetic field,
with the transverse component being expressed
in terms of the complex function
$
h_\perp ({\bf k}) 
= 
h_x ({\bf k}) + i h_y ({\bf k}).
$
In the limit where the internal $\mu_-$
and external $h_z$ Zeeman fields vanish and 
the spin-orbit field is null $(h_\perp = 0)$, the energies of the 
helicity bands are identical 
$\xi_{\Uparrow} ({\bf k})  = \xi_{\Downarrow} ({\bf k})$
producing no effect in the original energy 
dispersions~\cite{shenoy-2011}. 

When interactions are added to the problem, 
pairing can occur within the same helicity band 
(intra-helicity pairing) or between two different 
helicity bands (inter-helicity pairing). This
leads to a tensor order parameter for superfluidity that has
four components 
$\Delta_{\Uparrow \Uparrow} ({\bf k})
= 
-\Delta_T ({\bf k}) e^{-i\varphi},
$ 
corresponding to the helicity projection 
$\lambda = +1$;
$
\Delta_{\Uparrow \Downarrow} ({\bf k}) 
= 
- 
\Delta_S ({\bf k}),
$ 
and 
$
\Delta_{\Downarrow \Uparrow} ({\bf k}) 
= 
\Delta_S ({\bf k}),
$
corresponding to helicity projection $\lambda = 0$; and
$
\Delta_{\Downarrow \Downarrow} ({\bf k}) 
= 
-\Delta_T ({\bf k}) e^{i\varphi},
$
corresponding to helicity projection $\lambda = -1$.
The phase $\varphi ({\bf k})$ is defined from 
the amplitude-phase representation of the 
complex spin-orbit field 
$
h_{\perp} ({\bf k})
=  
\vert h_{\perp} ({\bf k}) \vert
e^{i\varphi ({\bf k})},
$
while the amplitude
$
\Delta_T ({\bf k}) 
= 
\Delta_0 
\vert h_\perp ({\bf k}) \vert
/
\vert {\bf h}_{\rm eff} ({\bf k}) \vert
$
for helicities $\lambda = \pm 1$ are directly 
proportional to the scalar order parameter 
$\Delta_0$ and to the relative magnitude of the spin-orbit
field $\vert h_\perp ({\bf k}) \vert$ with respect
to the magnitude of the effective magnetic field
$\vert {\bf h}_{\rm eff} ({\bf k}) \vert$. 
Additionally, $\Delta_T$ has the simple physical interpretation
of being the triplet component of the order parameter
in the helicity basis, which is induced by the presence
of non-zero spin-orbit field $h_\perp$, but vanishes when 
$h_\perp  = 0$.
Analogously the amplitude 
$
\Delta_S ({\bf k}) 
= 
\Delta_0 
h_\parallel ({\bf k})
/
\vert {\bf h}_{\rm eff} ({\bf k}) \vert
$
for helicity $\lambda = 0$ are directly 
proportional to the scalar order parameter 
$\Delta_0$ and to the relative magnitude of the total Zeeman
field $h_\parallel ({\bf k}) = \mu_- + h_z$ with respect
to the magnitude of the effective magnetic field
$\vert {\bf h}_{\rm eff} ({\bf k}) \vert$. 
Additionally, $\Delta_S$ has the simple physical interpretation
of being the singlet component of the order parameter
in the helicity basis. 
It is interesting to note the relation
$
\vert \Delta_T ({\bf k}) \vert^2
+
\vert \Delta_S ({\bf k}) \vert^2
=
\vert \Delta_0 \vert^2,
$
which, for fixed $\vert \Delta_0 \vert$, 
shows that as $\vert \Delta_S ({\bf k}) \vert$ increases,
$\vert \Delta_T ({\bf k}) \vert$ decreases and vice-versa. 
Such relation indicates that the singlet and 
triplet channels are not separable
in the presence of spin-orbit coupling.
Furthermore, the order parameter in the triplet sector 
$\Delta_{\Uparrow \Uparrow} ({\bf k}) $ and 
$\Delta_{\Downarrow \Downarrow} ({\bf k})$
contains not only $p$-wave, but also $f$-wave and even higher
odd angular momentum contributions, as long as 
the total Zeeman field $\mu_- + h_z$ is non-zero. 
Similarly, the order parameter in the singlet sector
$\Delta_{\Uparrow \Downarrow} ({\bf k}) $ and 
$\Delta_{\Downarrow \Uparrow} ({\bf k})$
contains not only only $s$-wave, but also $d$-wave and even
higher even angular momentum contributions, as long
as the total Zeeman field $\mu_- + h_z$ is non-zero. 
Higher angular momentum pairing in the helicity basis,
occurs because the original local (zero-ranged) interaction in the 
original $(\uparrow, \downarrow)$ 
spin basis is transformed into a finite-ranged
interaction in the helicity basis $(\Uparrow, \Downarrow)$.
In the limiting case of zero total Zeeman field
$\mu_- + h_z = 0$, the singlet component vanishes 
$( \Delta_S ({\bf k}) = 0 )$, while the triplet component
becomes independent of momentum $( \Delta_T({\bf k}) = \Delta_0 )$,
leading to order parameter 
$
\Delta_{\Uparrow \Uparrow} ({\bf k})
=
-h_{\perp}^* ({\bf k})
$, 
and
$
\Delta_{\Downarrow \Downarrow} ({\bf k})
= 
- h_{\perp} ({\bf k})
$
which contains only $p$-wave contributions~\cite{chuanwei-2008},
since the components of $h_{\perp} ({\bf k})$
depend linearly on momentum ${\bf k}$. 

The eigenvalues $E_j ({\bf k})$ of the Hamiltonian 
including the order parameter contribution emerge 
from the diagonalization of a $4\times 4$ 
matrix (see supplementary material). The two eigenvalues
for quasiparticles are 
\begin{equation}
\label{eqn:eigenvalue-1}
E_1 ({\bf k}) 
=
\sqrt{
\left(
\xi_{h-} 
-
\sqrt{
\xi_{h+}^2 
+
\vert \Delta_S ({\bf k}) \vert^2
}
\right)^2
+
\vert \Delta_T ({\bf k}) \vert^2
},
\end{equation}
corresponding to the highest-energy quasiparticle band, and
\begin{equation}
\label{eqn:eigenvalue-2}
E_2 ({\bf k}) 
=
\sqrt{
\left(
\xi_{h_-} 
+
\sqrt{
\xi_{h_+}^2 
+
\vert \Delta_S ({\bf k}) \vert^2
}
\right)^2
+
\vert \Delta_T ({\bf k}) \vert^2
},
\end{equation}
corresponding to the lowest-energy quasiparticle band,
while the eigenvalues for quasiholes are
$
E_3 ({\bf k}) 
= 
-
E_2 ({\bf k}) 
$
for highest-energy quasihole band
and
$
E_4 ({\bf k}) 
= 
-
E_1 ({\bf k})
$
for the lowest-energy quasihole band. 
The energy 
$
\xi_{h_-}
= 
\left[ 
\xi_\Uparrow ({\bf k}) 
- 
\xi_\Downarrow ({\bf k})
\right]
/
2
$
is momentum-dependent, corresponds to
the average energy difference between the
helicity bands and can be written as 
$
\xi_{h_-} 
=
- \vert {\bf h}_{\rm eff} ({\bf k}) \vert,
$
while the energy 
$
\xi_{h_+} 
= 
\left[ 
\xi_\Uparrow ({\bf k}) 
+ 
\xi_\Downarrow ({\bf k})
\right]
/
2
$
is also momentum dependent, corresponds
to the averaged energy sum of the helicity
bands and can be written as
$
\xi_{h_+} 
= 
K_+ ({\bf k}) = \epsilon_{\bf k} - \mu_+.
$

There are a few important points to notice about
the excitation spectrum of this system. First,
notice that $E_1 ({\bf k}) > E_2 ({\bf k}) \ge 0$.
Second, that the eigenergies are symmetric about zero, 
such that we can regard quasiholes (negative energy solutions) 
as anti-quasiparticles. Third, that only $E_2 ({\bf k})$
can have zeros (nodal regions) corresponding to the
locus in momentum space satisfying the following 
conditions:
a)
$
\xi_{h_-} 
= 
- \sqrt{
\xi_{h_+}^2 
+
\vert \Delta_S ({\bf k}) \vert^2
},
$
which corresponds physically 
to the equality between the effective magnetic field energy
$\vert {\bf h}_{\rm eff} ({\bf k}) \vert$ 
and the {\it excitation energy} for
the singlet component
$
\sqrt{
\xi_{h_+}^2 
+
\vert \Delta_S ({\bf k}) \vert^2
};
$  
and 
b) 
$
\vert
\Delta_T ({\bf k}) 
\vert
=
0,
$
corresponding to zeros 
of the triplet component of
the order parameter 
in momentum space.

Since $E_2 ({\bf k}) < E_1 ({\bf k})$, 
and only $E_2 ({\bf k})$ can have zeros, the low
energy physics is dominated by this eingenvalue.
In the case of equal Rashba-Dresselhaus (ERD)
where $h_\perp ({\bf k}) = v \vert k_x \vert$,
zeros of $E_2 ({\bf k})$ can occur when $k_x = 0$, 
leading to the following cases:
(a) two possible lines (rings) of nodes
at 
$
(k_y^2 + k_z^2)/(2m) 
= 
\mu_+ + \sqrt{(\mu_- + h_z)^2 - \vert \Delta_0 \vert^2}
$ 
for the outer ring,
and
$
(k_y^2 + k_z^2)/(2m) 
= 
\mu_+ - \sqrt{(\mu_- + h_z)^2 - \vert \Delta_0 \vert^2}
$ 
for the inner ring,
when $(\mu_- + h_z)^2 - \vert \Delta_0 \vert^2 > 0$;
(b) doubly-degenerate line of nodes 
$
(k_y^2 + k_z^2)/(2m) 
= 
\mu_+ 
$ 
for $\mu_+ > 0$,
doubly-degenerate point nodes
for $\mu_+ = 0$, or no-line of nodes
for $\mu_+ < 0$,
when 
$(\mu_- + h_z)^2 - \vert \Delta_0 \vert^2 = 0$;
(c) no line of nodes when $(\mu_- + h_z)^2 - \vert \Delta_0 \vert^2 < 0$.
In addition, case (a) can be refined into cases
(a2), (a1) and (a0). In case (a2), two rings indeed exist provided that 
$\mu_+ > \sqrt{(\mu_- + h_z)^2 - \vert \Delta_0 \vert^2}$.
However, the inner ring disappears when 
$\mu_+ = \sqrt{(\mu_- + h_z)^2 - \vert \Delta_0 \vert^2}$.
In case (a1), there is only one ring when
$
\vert 
\mu_+ 
\vert 
< 
\sqrt{(\mu_- + h_z)^2 - \vert \Delta_0 \vert^2},
$
In case (a0), the outer ring disappears at
$\mu_+ = - \sqrt{(\mu_- + h_z)^2 - \vert \Delta_0 \vert^2}$,
and for $\mu_+ < - \sqrt{(\mu_- + h_z)^2 - \vert \Delta_0 \vert^2}$
no rings exist. 

We choose our momentum, energy and velocity
scales through the Fermi momentum $k_{F+}$ defined from the 
total density of fermions
$
n_+ 
= 
n_\uparrow 
+ 
n_\downarrow
=
k_{F_+}^3/(3\pi^2). 
$ 
This choice leads to the
Fermi energy $\epsilon_{F_+} = k_{F_+}^2/2m$ 
and to the Fermi velocity $v_{F_+} = k_{F_+}/m$,
as energy and velocity scales respectively.
In Fig. 1, we show the phase diagram of 
Zeeman field $h_z/\epsilon_{F_+}$ versus
chemical potential $\mu_+/\epsilon_{F_+}$ 
describing possible superfluid phases according 
to their quasiparticle excitation spectrum. We label
the uniform superfluid phases with zero, one or two rings of nodes 
as US-0, US-1, and US-2, respectively.
Non-uniform (NU) phases also
emerge in regions where uniform phases are thermodynamically
unstable. The US-2/US-1 phase boundary is determined
by the condition $\mu_+ = \sqrt{(\mu_- + h_z)^2 - \vert \Delta_0 \vert^2}$,
when 
$\vert \mu_- + h_z \vert > \vert \Delta_0 \vert$;
the US-0/US-2 boundary is determined by
the Clogston-like condition $\vert (\mu_- + h_z) \vert = \vert \Delta_0 \vert$
when $\mu_+ > 0$, where the gapped US-0 phase  
disappears leading to the gapless US-2 phase;
and the US-0/US-1 phase boundary is determined by 
$\mu_+ = - \sqrt{(\mu_- + h_z)^2 - \vert \Delta_0 \vert^2}$,
when $\vert \mu_- + h_z \vert > \vert \Delta_0 \vert$.
Furthermore, with the US-0 boundaries, 
a crossover line between an indirectly gapped and a directly gapped 
US-0 phase occurs at $\mu_+ = 0$.
Lastly, some important multi-critical points
arise at the intersections of phase boundaries.
First the point $\mu_+ = 0$ and 
$\vert (\mu_- + h_z) \vert = \vert \Delta_0 \vert$ corresponds
to a tri-critical point for phases US-0, US-1, and US-2.
Second, the point $\vert \Delta_0 \vert = 0$
and $\mu_+  = \vert (\mu_- + h_z) \vert$ corresponds
to a tri-critical point for phases N, US-1 and US-2.
In the limit where both $\mu_-$ and $h_z$ vanish 
no phase transitions take place and the problem is reduced
to a crossover~\cite{zhai-2011, hu-2011, han-2011}.

\begin{figure} [htb]
\centering
\includegraphics[width=1.0\linewidth]{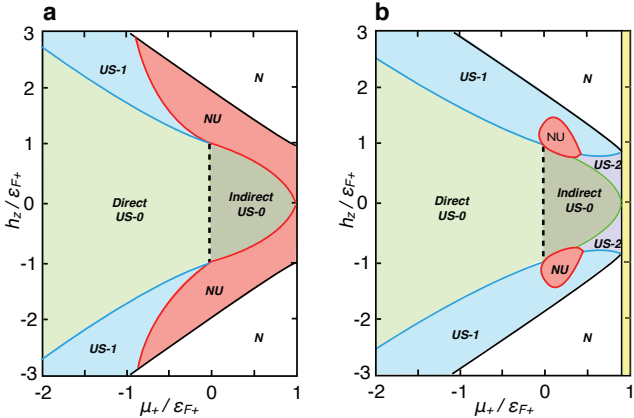}
\caption{ \label{fig:one} 
Phase diagram of Zeeman field $h_z/\epsilon_{F_+}$ versus 
chemical potential $\mu_+/\epsilon_{F_+}$ for 
a) $v/v_{F_+}= 0$ and b) $v/v_{F_+}= 0.28$ 
identifying uniform superfluid phases US-0 (gapped), US-1 (gapless with one
ring of nodes), and US-2 (gapless with two-rings of nodes).
The NU region corresponds to unstable uniform superfluids which
may include phase separation and/or a modulated superfluid (supersolid).
Solid lines represent phase boundaries, while the dashed line 
represents the crossover from the direct-gap to the indirect-gap US-0 phase.
}
\end{figure}

In the US-1 and US-2 phases near the zeros of 
$E_2 ({\bf k})$, quasiparticles have linear
dispersion and behave as Dirac fermions. Such change in
nodal structures is associated with bulk topological phase 
transitions of the Lifshitz class as noted for 
$p$-wave~\cite{volovik-1992} and $d$-wave~\cite{duncan-2000,botelho-2005} 
superfluids. Such Lifshitz topological phase transitions are possible here
because the spin-orbit coupling field induces the
triplet component of the order parameter $\Delta_T ({\bf k})$.
The loss of nodal regions correspond to annihilation 
of Dirac quasiparticles with opposite momenta, which 
lead to the disappearance of rings.
The transition from phase US-2 to indirect gapped US-0
occurs through the merger of the two-rings at the phase
boundary followed by the immediate opening of the indirect
gap at finite momentum. However, the transition from 
phase US-2 to US-1 corresponds to the disappearance of 
the inner ring through the origin of momenta, similarly 
the transition from US-1 to the directly gappped US-0 
corresponds to the disappearance of the last ring also through
the origin of momenta. In the case of Rashba-only coupling 
rings of nodes are absent and it is possible to have at most nodal points
~\cite{chuanwei-2011,iskin-2011}.
The last two phase transitions are special because 
the zero-momentum quasiparticles at these phase boundaries 
correspond to true Majorana zero 
energy modes if the phase $\varphi ({\bf k})$ of the
spin-orbit field 
$
h_\perp ({\bf k}) = 
\vert h_\perp ({\bf k}) \vert
e^{i \varphi ({\bf k})}
$ 
and the phase $\theta ({\bf k})$ 
of the order parameter 
$
\Delta_0 
= \vert \Delta_0 \vert 
e^{i \theta ({\bf k})}
$
have opposite phases at zero momentum:
$\varphi ({\bf 0}) = - \theta ({\bf 0})$ $[{\rm mod}(2\pi)]$. 
This can be seen from
an analysis of the quasiparticle eigenfunction 
$$
\Phi_2 ({\bf k}) 
= 
u_{1} ({\bf k}) \psi_{{\bf k} \uparrow} 
+
u_{2} ({\bf k}) \psi_{{\bf k} \downarrow} 
+
u_{3} ({\bf k}) \psi^\dagger_{-{\bf k} \uparrow} 
+
u_{4} ({\bf k}) \psi^\dagger_{-{\bf k} \downarrow} 
$$
corresponding to the eigenvalue $E_2 ({\bf k})$.
The emergence of zero-energy Majorana fermions requires the quasiparticle
to be its own anti-quasiparticle: $\Phi_2^\dagger ({\bf k}) = 
\Phi_2 ({\bf k})$. This can only happen at zero momentum 
${\bf k} = {\bf 0}$, where the amplitudes 
$u_1 ({\bf 0}) = u_3^* ({\bf 0})$ and  
$u_2 ({\bf 0}) = u_4^* ({\bf 0})$. Such requirement leads 
to the conditions $\mu_+^2 = (\mu_- + h_z)^2 + \vert \Delta_0 \vert^2$,
and $\varphi ({\bf 0}) = - \theta ({\bf 0})$ $[{\rm mod}(2\pi)]$, 
showing that Majorana fermions
can exist only at the US-0/US-1 and US-2/US-1 phase boundaries. It is 
important to emphasize that the Majorana fermions found here exist in the
bulk, and thus their emergence or disappeareance 
affect bulk thermodynamic properties, unlike
Majorana fermions found at the edge (surfaces) of topological 
insulators and some topological superfluids. The common ground between 
bulk and surface Majorana fermions is that both exist at boundaries: 
the bulk Majorana zero-energy modes may exist at the phase boundaries
between two topologically distinct superfluid phases, while 
surface Majorana zero-energy modes may exist at the spatial boundaries of
a topologically non-trivial superfluid.

It is evident that the transition between different superfluid phases
occurs without a change in symmetry in the order parameter $\Delta_0$,
and thus violates the symmetry-based Landau classification of phase 
transitions. In the present case, the simultaneous existence
of spin-orbit and Zeeman fields (internal or external) couple 
the singlet $\Delta_S ({\bf k})$ and triplet $\Delta_T ({\bf k})$ 
channels and all the superfluid phases US-0, US-1 and US-2 just have
different weights from each order parameter component. 
However a finer classification based on topological charges 
can be made via the construction of topological invariants. 
Since the superfluid phases US-0, US-1, US-2 are characterized 
by different excitation spectra corresponding
to the eigenvalues of the Hamiltonian matrix including 
interactions ${\bf H} ({\bf k})$, 
we can use the resolvent matrix 
$
{\bf R} (\omega, {\bf k})
= 
\left[
-\omega {\bf 1} +{\bf H} ({\bf k})
\right]^{-1}
$
and the methods of algebraic topology~\cite{nakahara-1990}
to construct the topological invariant
$$
\ell
=
\int_{\cal D}
\frac{dS_\gamma}{24\pi^2}
\epsilon^{\mu\nu\lambda\gamma}
{\rm Tr}
\left[
{\bf \Lambda}_{k_\mu} 
{\bf \Lambda}_{k_\nu} 
{\bf \Lambda}_{k_\lambda}
\right],
$$
where 
$
{\bf \Lambda}_{k_\mu} 
=
{\bf R}\partial_{k_\mu} {\bf R}^{-1}.
$
The topological invariant is $\ell = 0$
in the gapped US-0 phase,
is $\ell = 1$ in the gapless US-1 phase 
and $\ell  = 2$ in the gapless US-2 phase, 
showing that, for ERD spin-orbit coupling,  
$\ell$ counts the number of rings of zero-energy excitations
in each superfluid phase. 
The integral above has a hyper-surface measure $dS_\gamma$ 
and a domain ${\cal D}$ 
that encloses the region of zeros of $\omega =  E_j ({\bf k}) = 0$.
Here $\mu, \nu, \lambda, \gamma$ run from 0 to 3,
and $k_\mu$ has components $k_0 = \omega$, 
$k_1 = k_x$, $k_2 = k_y$, and $k_3 = k_z$.
The topological invariant measures
the flux of the four-dimensional vector 
$
F^\gamma =
\epsilon^{\mu\nu\lambda\gamma}
{\rm Tr}
\left[
{\bf \Lambda}_{k_\mu} 
{\bf \Lambda}_{k_\nu} 
{\bf \Lambda}_{k_\lambda}
\right]
/
{24\pi^2},
$
through a hypercube including the singular region
of the resolvent matrix ${\bf R} (\omega, {\bf k})$, much 
in the same way that the flux of the electric field ${\bf E}$ in 
Gauss' law of classical electromagnetism 
measures the electric charge $q$ enclosed by a Gaussian 
surface: $\oint d{\bf S} \cdot (\epsilon_0 {\bf E}) = q$.
Thus, the topological invariant
defined above defines the topological charge of 
fermionic excitations, in the same sense as Gauss' law for 
the electric flux defines the electric charge.

\begin{figure} [htb]
\centering
\includegraphics[width=1.0\linewidth]{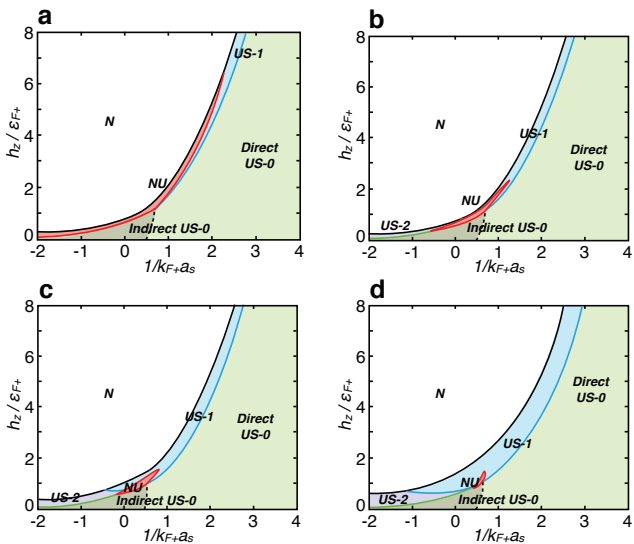}
\caption{ \label{fig:two} 
Phase diagram of Zeeman field $h_z/\epsilon_{F_+}$ 
versus interaction $1/(k_{F+} a_s)$ showing
uniform superfluid phases US-0, US-1, and US-2,
and non-uniform (NU) region for 
a) $v/v_{F_+} = 0$; b) $v/v_{F_+} = 0.14$;
c) $v/v_{F_+} = 0.28$; d) $v/v_{F_+} = 0.56$.
Solid lines are phase boundaries, 
the dashed line indicates a crossover from the
indirect- to direct-gapped US-0.
}
\end{figure}

A full phase diagram can be constructed only upon verification
of thermodynamic stability of all the proposed phases. For this
purpose it becomes imperative to investigate the maximum 
entropy condition (see supplementary material). Independent of any microscopic
approximations, the necessary and sufficient
conditions for thermodynamic stability of a given phase are:
positive isovolumetric heat capacity
$
C_V 
= 
T 
\left(
\partial S
/
\partial T
\right)_{V, \{N_\alpha \} }
\ge 0;
$
positive
chemical susceptibility matrix 
$
\xi_{\alpha \beta} 
= 
\left(
\partial \mu_\alpha / \partial N_\beta
\right)_{T, V},
$
i.e, eigenvalues of the matrix $\left[ \xi \right]$ are both positive;
and positive bulk modulus 
$
B 
= 
1/\kappa_T 
$
or isothermal 
compressibility 
$
\kappa_T 
= 
- V^{-1} 
\left(
\partial V /\partial P
\right)_{T, \{ N_\alpha \} }.
$
Using these conditions, we construct the full phase diagram 
described in Fig.~\ref{fig:one} for equal Rasha-Dresselhaus
(ERD) spin-orbit coupling. The regions, where the uniform superfluid
phases are unstable are labeled by the abbreviation NU to indicate 
that non-uniform phases such as phase separation or modulated 
superfluid (supersolid) may emerge.  In Fig.~\ref{fig:two}, we show the phase
diagram of Zeeman field $h_z/\epsilon_{F_+}$ versus interaction parameter
$1/(k_{F+} a_s)$, for population balanced fermions, where
the number of spin-up fermions $N_\uparrow$ is equal
to the number of spin-down fermions $N_\downarrow$.

\begin{figure} [htb]
\centering
\includegraphics[width=1.0\linewidth]{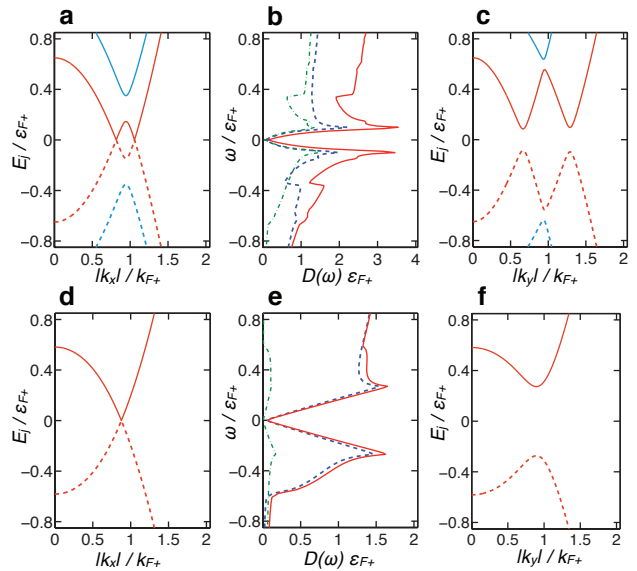}
\caption{ \label{fig:three} 
Energy spectrum and density of states in phase US-2 are 
shown in a), b), c) for $h_z/\epsilon_{F_+} = 0.5$
and $1/(k_{F_+}a_s) = -0.4$ and in phase US-1 are
shown in d), e), f) for $h_z/\epsilon_{F_+} = 2.0$
and $1/(k_{F_+}a_s) = 1.0$. Energies $E_j (k_x, 0, 0)$ versus 
$\vert k_x \vert$ in a) and d); frequency $\omega$ versus density of states 
${\cal D}_{\uparrow} (\omega)$ (dashed), 
${\cal D}_{\downarrow} (\omega)$ (dot-dashed),
and their sum ${\cal D} (\omega)$ (solid) in b) and e);  
energies $E_j (0, k_y, 0)$ versus $\vert k_y \vert$ in c) and f).
}
\end{figure}

Since these superfluid phases exhibit major changes in 
momentum-frequency space as evidenced by their single particle
excitation spectrum, it is important to explore additional
spectroscopic quantitities to characterize further the 
nature of these phases and the phase transitions between them.
An important quantity is the $4\times 4$ resolvent matrix 
\begin{equation}
\label{eqn:resolvent-matrix}
{\bf R}(i\omega, {\bf k})
=
\left(
\begin{array}{cc}
{\bf G} (i\omega, {\bf k}) & {\bf F} (i\omega, {\bf k}) \\
{\bf F}^\dagger (i\omega, {\bf k}) & {\overline {\bf G}} (i\omega, {\bf k}) \\
\end{array}
\right),
\end{equation}
from where the spectral density 
$
{\cal A}_{\alpha} (\omega, {\bf k}) 
= 
-(1/\pi) 
{\rm Im} G_{\alpha \alpha} (i\omega = \omega + i\delta,
{\bf k})
$
for spin $\alpha = \uparrow, \downarrow$ can be extracted.
The spectral function ${\cal A}_{\alpha} (\omega, {\bf k})$ 
in the plane of momenta $k_y$-$k_z$ with $k_x = 0$ and 
frequency $\omega = 0$ reveals the existence
of rings of zero-energy excitations in the US-1 and US-2 phases.
The density of states 
$
{\cal D}_{\alpha} (\omega) 
= 
\sum_{\bf k} {\cal A}_{\alpha} (\omega, {\bf k})
$ 
for spin $\alpha$ as a function of frequency $\omega$
is also an important spectroscopic quantity which
is shown in Fig.~\ref{fig:three} along with 
excitation spectra $E_j({\bf k})$ for phases US-1 and
US-2 at fixed ERD spin-orbit coupling $v/v_{F_+} = 0.28$.
The parameters used for phase US-1 are 
$h_z/\epsilon_{F_+} = 0.5$ and $1/(k_{F_+} a_s) = -0.4$,
while for phase US-2 they are
$h_z/\epsilon_{F_+} = 2.0$ and $1/(k_{F_+} a_s) = 1.0$.
Notice that, even though the excitation spectrum $E_j ({\bf k})$ is symmetric,
the coherence factors appearing in the matrix ${\bf G}$ are not, 
such that the density of states ${\cal D}_{\alpha} (\omega)$ 
is not an even function of $\omega$, and
thus it is not particle-hole symmetric. The main feature of
${\cal D}_{\alpha} (\omega)$ at low frequencies is the linear behavior
due to the existence of Dirac quasiparticles and quasiholes in the
US-1 and US-2 phases, which are absent in the direct-gap and 
the indirect-gap US-0 phases. The peaks and structures in 
${\cal D}_{\alpha} (\omega)$ mostly emerge due to 
the maxima and minima of $E_j ({\bf k})$. Notice that for 
finite Zeeman field $h_z$, the density of states 
${\cal D}_{\uparrow} (\omega) 
\ne {\cal D}_{\downarrow} (\omega)$
because the induced population imbalance 
$P = (N_{\uparrow} - N_{\downarrow})/(N_{\uparrow} + N_{\downarrow})$
is non-zero. For the US-2 case shown in Fig.~\ref{fig:three}b, the 
induced population imbalance $P \ll 1$ since $h_z/\epsilon_{F_+}$
is small, while for the US-1 case shown in Fig.~\ref{fig:three}e, 
$P \approx 1$ as the spins are almost fully polarized since
$h_z/\epsilon_{F_+}$ is large.

%

%
%
In summary, we have discussed the effects of spin-orbit and Zeeman
fields in ultra-cold Fermi superfluids, obtained the phase diagrams of
Zeeman field versus interaction parameter or versus chemical potential, 
and identified several bulk topological phase transitions between 
gapped and gapless superfluids as well as a variety of multi-critical
points. We have shown that the presence of simultaneous Zeeman and
spin-orbit fields induces higher angular momentum pairing, as manifested
in the emergence of momentum dependence of the singlet and triplet 
components of order parameter expressed in the helicity basis. 
Finally, we have characterized topological phases and phase
transitions between them through their excitation spectra
(existence of Dirac quasiparticles or Majorana zero-energy modes),
topological charges, and spectroscopic and 
thermodynamic properties, such as density of states and 
isothermal compressibility.

\acknowledgements{We thank ARO (W911NF-09-1-0220) for support.}

\section{Artificial spin-orbit coupling in ultra-cold Fermi superfluids: 
(Supplementary Material)}

%
%

The method used to study the spin-orbit and Zeeman effects in ultra-cold
Fermi superfluids is the functional integral method and its saddle-point
approximation in conjunction with fluctuation effects. To describe the 
thermodynamic phases and the corresponding phase diagram in terms of the
interactions, Zeeman and spin-orbit fields, we calculate partition
function at temperature $T$ 
$
Z = \int \mathcal{D}[\psi, \psi^\dagger] \exp \left(
 -S[\psi, \psi^\dagger]
\right) 
$ 
with action
$$
\label{eqn:action-initial}
S[\psi, \psi^\dagger]
=
\int d\tau d {\bf r}
\left[
\sum_{\alpha}
\psi^\dagger_{\alpha} ({\bf r}, \tau)
\frac{\partial}{\partial \tau}
\psi_{\alpha} ({\bf r}, \tau) +
{\cal H} ({\bf r}, \tau)
\right],
$$
where the Hamiltonian density is given in Eq.~(\ref{eqn:hamiltonian}).

Using the standard Hubbard-Stratanovich transformation that
introduces the pairing field
$
\Delta ({\bf r}, \tau)
=
g
\langle
\psi_{\downarrow} ({\bf r}, \tau)
\psi_{\uparrow} ({\bf r}, \tau)
\rangle
$
and integrating over the fermion variables lead to the effective 
action
$$
S_{\rm eff}
=
\int d\tau d {\bf r}
\left[
\frac{ \vert \Delta ({\bf r, \tau}) \vert^2 }
{g}
-
\frac{T}{2V}
\ln \det \frac{{\bf M}}{T}
+
\widetilde K_+ \delta ({\bf r} - {\bf r}^\prime)
\right],
$$
where
$
\widetilde K_{+}
=
( \widetilde K_\uparrow + \widetilde K_\downarrow )/2.
$
The matrix ${\bf M}$ is
\begin{equation}
\label{eqn:matrix-m}
{\bf M}
=
\left(
\begin{array}{cccc}
\partial_\tau + \widetilde K_\uparrow & - h_\perp & 0 & -\Delta \\
- h_\perp^* & \partial_\tau + \widetilde K_\downarrow &  \Delta & 0 \\
0 & \Delta^\dagger & \partial_\tau - \widetilde K_\uparrow &  h_\perp^* \\
-\Delta^\dagger  & 0 & h_\perp & \partial_\tau - \widetilde K_\downarrow
\end{array}
\right),
\end{equation}
where $h_{\perp} = h_x - i h_y$ corresponds to the transverse component
of the spin-orbit field, $h_z$ to the parallel component
with respect to the quantization axis $z$,
$\widetilde K_\uparrow  = {\hat K}_\uparrow - h_z$,
and $\widetilde K_\downarrow  = {\hat K}_\downarrow + h_z$.

%
%

To make progress, we use the saddle point
approximation $\Delta ({\bf r}, \tau) = \Delta_0 + \eta ({\bf r}, \tau),$ 
and write ${\bf M} = {\bf M}_{\rm sp} + {\bf M}_{\rm f}$.
The matrix ${\bf M}_{\rm sp}$ is obtained via the saddle point
$\Delta ({\bf r}, \tau) \to \Delta_0$ which takes 
${\bf M} \to {\bf M}_{\rm sp}$,
and the fluctuation matrix ${\bf M}_{{\rm f}} = {\bf M} - {\bf M_{\rm sp}}$
depends only on $\eta ({\bf r}, \tau)$ and its Hermitian conjugate.
Thus, we write the effective action as
$S_{\rm eff} = S_{\rm sp} + S_{\rm f}$. The first term is 
$$
S_{\rm sp}
=
\frac{V}{T}
\frac{\vert \Delta_0 \vert^2}{g}
-\frac{1}{2}
\sum_{{\bf k}, i\omega_n, j}
\ln
\left[
\frac{- i\omega_n + E_j ({\bf k})}{T}
\right]
+
\sum_{\bf k}
\frac{{\widetilde K}_{+}}{T},
$$
in momentum-frequency coordinates $({\bf k}, i\omega_n)$, 
where $\omega_n = (2n + 1) \pi T$. Here, $E_j ({\bf k})$ 
are the eigenvalues of
\begin{equation}
{\bf H}_{\rm sp}
=
\left(
\begin{array}{cccc}
\widetilde K_\uparrow ({\bf k}) & - h_\perp ({\bf k}) & 0 & -\Delta_0 \\
- h_\perp^* ({\bf k}) & \widetilde K_\downarrow ({\bf k}) &  \Delta_0 & 0 \\
0 & \Delta_0^\dagger & -\widetilde K_\uparrow ({-\bf k}) &  h_\perp^* ({-\bf k}) \\
-\Delta_0^\dagger  & 0 & h_\perp (-{\bf k}) & - \widetilde K_\downarrow (-{\bf k})
\end{array}
\right),
\end{equation}
which describes the Hamiltonian of elementary excitations in the
four-dimensional basis 
$ 
\Psi^\dagger 
= 
\left\{
\psi_{\uparrow}^\dagger ({\bf k}), 
\psi_{\downarrow}^\dagger ({\bf k}), 
\psi_{\uparrow}(-{\bf k}), 
\psi_{\downarrow}(-{\bf k})
\right\}. 
$ 
The fluctuation action is 
$$
S_{\rm f}
=
\int d\tau d{\bf r}
\left[
\frac{\vert \eta ({\bf r}, \tau) \vert^2}{g}
-
\frac{T}{2V}
\ln \det
\left(
{\bf 1} + {\bf M}_{\rm sp}^{-1} {\bf M}_{\rm f}
\right)
\right].
$$
The spin-orbit field is 
$ 
{\bf h}_\perp ({\bf k}) = 
{\bf h}_R ({\bf k}) + {\bf h}_D ({\bf k})$,
where  
$
{\bf h}_R ({\bf k}) 
= 
v_R \left( -k_y {\hat{\bf x}} + k_x {\hat {\bf y}} \right)$ 
is of Rashba-type
and
$ {\bf h}_D ({\bf k}) = v_D \left( k_y {\hat {\bf
x}} + k_x {\hat {\bf y}} \right) $ 
is of Dresselhaus-type,
has magnitude 
$ 
\vert h_{\perp} ({\bf k}) \vert
= 
\sqrt{ \left( v_D -
v_R \right)^2 k_y^2 + \left( v_D + v_R \right)^2 k_x^2 }. 
$ 
For Rashba-only (RO) $(v_D = 0)$ and for equal
Rashba-Dresselhaus (ERD) couplings $(v_R = v_D = v/2)$, 
the magnitude of the transverse fields are 
$ 
\vert h_{\perp} ({\bf k}) \vert 
= 
v_R \sqrt{k_x^2 + k_y^2}
$ 
($v_R > 0$) and 
$ 
h_{\perp} ({\bf k}) = v \vert k_x \vert 
$
($v > 0$), respectively.   

The Hamiltonian in the helicity basis 
$\Phi = {\bf U} \Psi$, where ${\bf U}$ is the unitary matrix
that diagonalizes the Hamiltonian in the normal state, is  
$$
\widetilde{\bf H}_{\rm sp} ({\bf k})
=
\left(
\begin{array}{cccc}
\xi_{\Uparrow}({\bf k}) & 0 & 
\Delta_{\Uparrow \Uparrow} ({\bf k}) & \Delta_{\Uparrow \Downarrow} ({\bf k}) \\
0  & \xi_{\Downarrow}({\bf k})&  
\Delta_{\Downarrow \Uparrow} ({\bf k}) & \Delta_{\Downarrow \Downarrow} ({\bf k}) \\
\Delta_{\Uparrow \Uparrow}^* ({\bf k}) & \Delta_{\Uparrow \Downarrow}^* ({\bf k})& 
- \xi_{\Uparrow}({\bf k}) &  0 \\
\Delta_{\Uparrow \Downarrow}^* ({\bf k}) & \Delta_{\Downarrow\Downarrow}^* ({\bf k}) & 
0  & -\xi_{\Downarrow}({\bf k})
\end{array}
\right).
$$
The components of the order parameter 
in the helicity basis are given by 
$
\Delta_{\Uparrow \Uparrow} ({\bf k}) 
=  
\Delta_T ({\bf k}) e^{-i\varphi_{\bf k}},
$
and 
$
\Delta_{\Downarrow \Downarrow} ({\bf k}) =  
-\Delta_T ({\bf k}) e^{i\varphi_{\bf k}}
$
for the triplet channel and by 
$
\Delta_{\Uparrow \Downarrow} ({\bf k}) =  -\Delta_S ({\bf k})
$
and 
$
\Delta_{\Downarrow \Uparrow} ({\bf k}) =  \Delta_S ({\bf k})
$
for the singlet channel.
The eigenvalues of ${\bf H}_{sp} ({\bf k})$ for quasiparticles
$E_1 ({\bf k})$, $E_2 ({\bf k})$ are listed in Eqs. 
(\ref{eqn:eigenvalue-1}) and (\ref{eqn:eigenvalue-2}),
while the eigenvalues for quasiholes are
$E_3 ({\bf k}) = - E_2 ({\bf k})$, 
and $E_4 ({\bf k}) = - E_1 ({\bf k})$.

The thermodynamic potential is
$\Omega = \Omega_{\rm sp} + \Omega_{\rm f}$,
where 
$$
\Omega_{\rm sp}
=
V
\frac{\vert \Delta_0 \vert^2}{g}
-\frac{T}{2}
\sum_{{\bf k}, j}
\ln
\left\{
1 + \exp \left[ - E_j ({\bf k})/T \right]
\right\}
+
\sum_{\bf k}
{\bar K}_{+},
$$
with
$
{\bar K}_{+}
=
\left[
\widetilde K_\uparrow (-{\bf k})
+
\widetilde K_\downarrow (-{\bf k})
\right]/2
$
is the saddle point contribution and 
$\Omega_{\rm f} = - T \ln Z_{\rm f}$, with
$
Z_{\rm f} = 
\int {\cal D} [{\bar \eta}, {\eta}]
\exp 
\left[
- S_{\rm f} ({\bar \eta}, {\eta}) 
\right]
$
is the fluctuation contribution.
The order parameter is determined via the minimization of $\Omega_{\rm sp}$
with respect to $\vert \Delta_0 \vert^2$, leading to
\begin{equation}
\label{eqn:order-parameter-general} 
\frac{V}{g} 
= 
-\frac{1}{2}
\sum_{{\bf k}, j} 
n_F \left[E_j ({\bf k}) \right] 
\frac{\partial E_j ({\bf k})}{\partial \vert \Delta_0 \vert^2},
\end{equation}
where
$ 
n_F \left[  E_j (\mathbf{k})  \right] 
= 
1/(\exp\left[ E_j ({\bf k})/T\right] + 1) 
$ 
is the Fermi function for energy $E_j ({\bf k})$. 
The contact interaction $g$ is expressed in terms
of the scattering parameter $a_s$ via the Lippman-Schwinger relation
discussed in the main text.

The total number of particles $N_+ = N_\uparrow + N_\downarrow$ is
defined from the thermodynamic relation
$
N_{+} = -
\left(
\partial \Omega
/
\partial \mu_{+}
\right)_{T, V},
$
and can be written as 
\begin{equation}
\label{eqn:number-general}
N_{+} 
= 
N_{\rm sp} + N_{\rm f}.
\end{equation}
The saddle point contribution is 
$$
N_{\rm sp}
=
- 
\left(
\frac{\partial \Omega_{\rm sp}}{ \partial \mu_{+} }
\right)_{T,V}
=
\frac{1}{2}
\sum_{\bf k}
\left[
1 -
\sum_j n_F \left[ E_j ({\bf k}) \right]
\frac{\partial E_j ({\bf k})}{\partial \mu_{+}}
\right],
$$
and the fluctuation contribution is
$
N_{\rm f}
= 
- 
\left(
\partial \Omega_{\rm f}
/
\partial \mu_{+},
\right)_{T,V}
$
leading to 
$$
N_{\rm f} 
=
\frac{T}{Z_{\rm f}} 
\int {\cal D} 
\left[
{\bar \eta}, \eta
\right]
\exp 
\left[
-S_{\rm f} ({\bar \eta}, \eta)
\right]
\left(
-
\frac{\partial S_{\rm f}({\bar \eta}, \eta)}{\partial \mu_+}
\right),
$$
with the partial derivative being
$$
\frac{\partial S_F ({\bar \eta}, \eta)}{\partial \mu_+}
=
-\frac{T}{2V}
{\rm Tr}
\left[
\left(
 1 + {\bf M}_{\rm sp}^{-1} {\bf M}_{\rm f}
\right)^{-1}
\frac{\partial}{\partial \mu_+}
\left(
{\bf M}_{\rm sp}^{-1} {\bf M}_{\rm f}
\right)
\right].
$$

\begin{figure} [htb]
\centering
\includegraphics[width=1.0\linewidth]{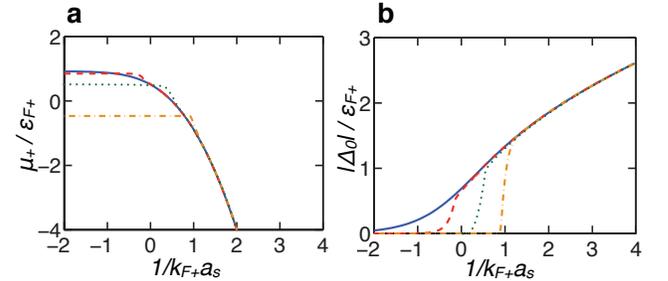}
\caption{ \label{fig:four} 
a) Chemical potential $\mu_+/\epsilon_{F_+}$ and
b) order parameter amplitude $\vert \Delta_0 \vert/\epsilon_{F_+}$ 
versus interaction parameter $1/(k_{F_+} a_s)$
for spin-orbit parameter $v/v_{F_+} = 0.28$ and
values of the Zeeman field
$h_z/\epsilon_{F_+} = 0$ (solid); 
$h_z/\epsilon_{F_+} = 0.5$ (dashed);
$h_z/\epsilon_{F_+} = 1.0$ (dotted);
and 
$h_z/\epsilon_{F_+} = 2.0$ (dot-dashed).
}
\end{figure}

Knowledge of the thermodynamic potential $\Omega$, of the order parameter
Eq.~(\ref{eqn:order-parameter-general}) and number 
Eq.~(\ref{eqn:number-general}) provides a complete theory for 
spectroscopic and thermodynamic properties of attractive ultra-cold fermions
in the presence of Zeeman and spin-orbit fields.
Representative Saddle point solutions for chemical potential $\mu_+$ and
order parameter amplitude $\vert \Delta_0 \vert$ 
as a function of $1/(k_{F+}a_s)$ in the equal Rashba-Dresselhaus (ERD) 
case $(v/v_{F_+} = 0.28)$ are shown 
in Fig.~\ref{fig:four} 
for $h_z/\epsilon_{F+} = 0, 0.5, 1.0, 2.0$.
These parameters are used to obtain the phase diagrams described 
in Figs.~\ref{fig:one} and~\ref{fig:two} in combination with
an analysis of the excitation spectrum $E_j ({\bf k})$ 
given in Eqs.~(\ref{eqn:eigenvalue-1}) and~(\ref{eqn:eigenvalue-2})
and the thermodynamic stability conditions for all the uniform 
superfluid phases: directly or indirectly
gapped superfluid with zero nodal rings (US-0); gapless superfluid
with one ring of nodes (US-1); and gapless superfluid
with two rings of nodes (US-2).

A thermodynamic stability analysis of all proposed phases can 
be performed by investigating the maximum 
entropy condition. The total change in entropy due to thermodynamic
fluctuations, irrespective to any approximations imposed on the
microscopic Hamiltonian, can be written as 
$$
\Delta S_{\rm tot} 
= 
- 
\frac{1}{2T}
\left(
\Delta T \Delta S
- 
\Delta P \Delta V
+ 
\Delta \mu_\alpha \Delta N_\alpha
\right),
$$
where the repeated $\alpha$ index indicates summation,
and the condition $\Delta S_{\rm tot} \le 0$ guarantees that the
entropy is maximum.
Considering the entropy $S$ to be a function of
temperature $T$, number of particles $N_\alpha$ and
volume $V$, we can elliminate the fluctuations $\Delta S$,
$\Delta P$, and $\Delta \mu_{\alpha}$ in favor of fluctuations
$\Delta T$, $\Delta V$ and $\Delta N_{\alpha}$, and show
that the fluctuations $\Delta T$ are statistically independent
of $\Delta N_{\alpha}$ and $\Delta V$, while fluctuations 
$\Delta N_{\alpha}$ and $\Delta V$ are not. The first condition
for thermodynamic stability 
leads to the requirement that the isovolumetric 
heat capacity
$
C_V 
= 
T 
\left(
\partial S
/
\partial T
\right)_{V, \{N_\alpha \} }
\ge 0.
$
Additional conditions are directly related to number
$\Delta N_\alpha$ and volume $\Delta V$ fluctuations.
They require the  
chemical susceptibility matrix 
$
\xi_{\alpha \beta} 
= 
\left(
\partial \mu_\alpha / \partial N_\beta
\right)_{T, V}
$
to be positive definite, i.e,
that its eigenvalues are both positive.
This is guaranteed by 
$
\rm {det} [ \xi ] 
= 
\xi_{\uparrow \uparrow} \xi_{\downarrow \downarrow}
- 
\xi_{\uparrow \downarrow} \xi_{\downarrow \uparrow}
> 0
$
and $\xi_{\uparrow \uparrow} > 0$.
The last condition for thermodynamic stability is that the bulk modulus 
$
B 
= 
1/\kappa_T 
$
or the isothermal 
compressibility 
$
\kappa_T 
= 
- V^{-1} 
\left(
\partial V /\partial P
\right)_{T, \{ N_\alpha \} }, 
$
are positive. 
Since the number $\Delta N_{\alpha}$ and volume $\Delta V$ 
fluctuations are not statistically independent, the bulk modulus
is related to the matrix $\left[ \xi \right]$
via 
$
V/\kappa_T  
=
N_\uparrow^2 \xi_{\uparrow \uparrow}
+ 
N_\uparrow N_\downarrow \xi_{\uparrow \downarrow}
+
N_\downarrow N_\uparrow  \xi_{\downarrow \uparrow} 
+
N_\downarrow^2 \xi_{\downarrow \downarrow}.
$
The positivity of the volumetric specific heat $C_V$, chemical
susceptibility matrix $\left[ \xi \right]$ and bulk modulus
$B = 1/\kappa_T$ are the necessary and sufficient conditions for
thermodynamic stability, which must be satisfied 
irrespective of approximations used at the microscopic level.

\begin{figure} [htb]
\centering
\includegraphics[width=1.0\linewidth]{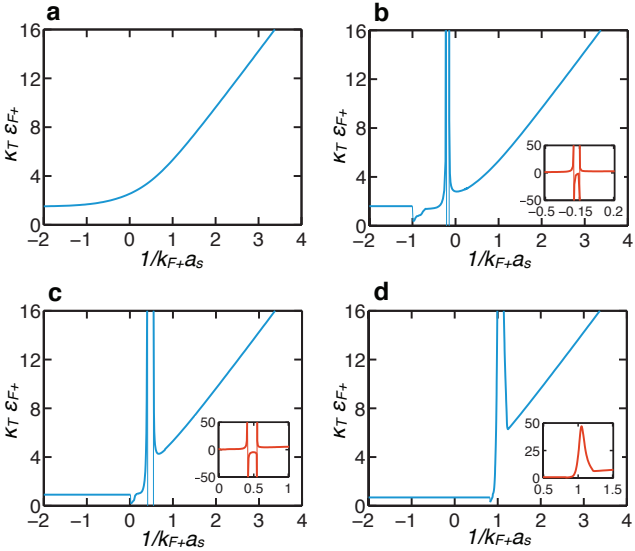}
\caption{ \label{fig:five} 
Isothermal compressibility 
${\bar \kappa}_T 
= (N_+^2)V^{-1} \kappa_T = 
\left(
\partial N_+ / \partial \mu_+
\right)_{T,V}
$
in units of $3 N_+/(4\epsilon_{F_+})$
versus interaction $1/(k_{F+} a_s)$ 
at spin-orbit coupling $v/v_{{F_+}} = 0.28$ 
for the values of the Zeeman field
a) $h_z/\epsilon_{F+} = 0$; 
b) $h_z/\epsilon_{F+} = 0.5$;
c) $h_z/\epsilon_{F+} = 1.0$;
and d) $h_z/\epsilon_{F+} = 2.0$.
Insets show regions where the compressibility is large.
}
\end{figure}

Further characterization of phases US-0, US-1 and US-2 
is made via thermodynamic properties such as the
isothermal compressibility
$
\kappa_T = (V/N_+^2)
\left(
\partial N_+/\partial \mu_+ 
\right)_{T, V},
$ 
which is shown in Fig.~\ref{fig:five} versus
$1/(k_{F+} a_s)$ for the values of the Zeeman field
$h_z/\epsilon_{F+} = 0, 0.5, 1.0, 2.0$ and spin-orbit
coupling $v/v_{F+} = 0.28$. Notice the negative regions
of $\kappa_T$ indicating that the uniform superfluid
phases are unstable, and its discontinuities at
phase boundaries.
The normal state compressibility $\kappa_T$ 
or 
$
{\bar \kappa}_T 
= (N_+^2)V^{-1} \kappa_T = 
\left(
\partial N_+ / \partial \mu_+
\right)_{T,V}
$
can be obtained analytically
for arbitrary Zeeman $h_z$ and spin-orbit parameter $v$
in the BCS limit where $1/k_{F_+} a_s \to - \infty$ as
\begin{equation}
{\bar \kappa}_T
=
\frac{3N_+}{4\epsilon_{F+}}\sum_{j=\pm}
\left[ A_j +
\left[
\tilde \mu_+ - A_j^2 + \sqrt{\tilde h_z^2+2\tilde v A_j^2}
\right]
\frac{\partial A_j}{\partial \tilde \mu_+}
\right],
\end{equation}
where the auxiliary function $A_j$ is
$$
A_\pm =
\sqrt{
\left(\tilde \mu_+ + \tilde v\right)
\pm
\sqrt{
\left( \tilde \mu_+ + \tilde v\right)^2
-\left(\tilde \mu_+^2-\tilde h_z^2\right)
}}
$$
and its derivative is
$$
\frac{\partial A_{\pm}}{\partial \tilde \mu_+}
=
\left[ 
1 \pm \tilde v / \sqrt{
( \tilde \mu_+ + \tilde v)^2-(\tilde \mu_+^2-\tilde h_z^2)
}
\right]/ (2A_{\pm})
$$
with $\tilde \mu_+ = \mu_+ / \epsilon_{F+}$, 
$\tilde h_z = h_z / \epsilon_{F+}$, and $\tilde v = v / (2\epsilon_{F+})$.
Notice that, as $h_z \to 0$ and $\tilde v \to 0$, 
$A_\pm \to \sqrt{\tilde \mu_+}$ and 
${\bar \kappa}_T \to (3N_+)/(2\epsilon_{F+})$ is reduced to the 
standard result, since $\tilde \mu_+ \to 1$. 
In addition, $\kappa_T$ or ${\bar \kappa}_T$ can be
obtained analytically in the BEC limit where 
$1/k_{F_+} a_s \to + \infty$. When $h_z$ and $v$ are zero,
then
\begin{equation}
{\bar \kappa}_T 
= 
\frac{3 N_+}{2\epsilon_{F_+}}
\frac{\pi}{k_{F_+} a_s }
\end{equation}
can also be written in terms of bosonic properties
\begin{equation}
\frac{1}{V}
\left(
\frac{\partial N_+}{\partial \mu_+}
\right)_{T,V}
=
\frac{1}{\pi} 
\left(
\frac{m_B}{a_B}
\right),
\end{equation}
where 
$m_B = 2 m$ is the boson mass and $a_B = 2 a_s$
in the boson-boson interaction.
In the case where $h_z \ne 0$ and $v \ne 0$, a 
similar expression can be derived for 
$
V^{-1} 
\left(
\partial N_+
/
\partial \mu_+
\right)_{T,V}
$
but the effective boson mass $m_B = 2m f(h_z, v)$, 
and the effective boson-boson interaction 
$a_B = 2a_s g (h_z, v)$ are now functions of $h_z$ and $v$.
Notice that the ratio $m_B/a_B$ in the BEC limit 
can be directly extracted from the behavior of
$\bar \kappa_T$ for large $1/(k_{F_+} a_s)$.


\begin{thebibliography}{99}

%
\bibitem{spielman-2011}
Y. J. Lin, K. Jimenez-Garcia, and I. B. Spielman,
Nature {\bf 471}, 83-86 (2011).

%
\bibitem{sinova-2009}
X. J. Liu, M. F. Borunda, X. Liu, and J. Sinova,
Phys. Rev. Lett. {\bf 102}, 046402 (2009).

%
\bibitem{chapman-sademelo-2011}
M. Chapman and C. S{\'a} de Melo, 
Nature {\bf 471}, 41-42 (2011).

%
\bibitem{kane-2005}
C. L. Kane and E. J. Mele,
Phys. Rev. Lett. {\bf 95}, 146802 (2005).

%
\bibitem{gorkov-2001}
L. P. Gor'kov and E. I. Rashba,
Phys. Rev. Lett. {\bf 87}, 037004 (2001).

%
\bibitem{galitski-2007}
T. D. Stanescu, C. Zhang, and V. Galitski,
Phys. Rev. Lett. {\bf 99} 110403 (2007).

%
\bibitem{dresselhaus-1955}
G. Dresselhaus,
Phys. Rev. {\bf 100}, 580 (1955).

%
\bibitem{rashba-1984}
Y. A. Bychkov and E. I. Rashba, 
J. Phys. C {\bf 17}, 6039 (1984).

%
\bibitem{dalibard-2010}
J. Dalibard, F. Gerbier, G. Juzeliunas, and P. Ohberg, 
arXiv:1008.5378 (2010).

%
\bibitem{shenoy-2011}
J. P. Vyasanakere, S. Zhang, and V. B. Shenoy, 
arXiv:1104.5633 (2011).

%
\bibitem{chuanwei-2008}
C. Zhang, S. Tewari, R. M. Lutchyn, and S. Das Sarma, 
Phys. Rev. Lett. {\bf 101} 160401 (2008).

%
\bibitem{zhai-2011}
Z.-Q. Yu and H. Zhai, 
arXiv:1105.2250 (2011).

%
\bibitem{hu-2011}
H. Hu, L. Jiang, X. Jiu, and H. Pu, 
arXiv:1105.2488 (2011).

%
\bibitem{han-2011}
L. Han and C. A. R. S\'a de Melo, 
arXiv:1106.3613 (2011).

%
\bibitem{volovik-1992}
G. E. Volovik, 
World Scientific, Singapore (1992).

%
\bibitem{duncan-2000}
R. D. Duncan and C. A. R. S\'a de Melo, 
Phys. Rev. B {\bf 62}, 9675 (2000).

%
\bibitem{botelho-2005}
S. S. Botelho and C. A. R. S\'a de Melo, 
Phys. Rev. B {\bf 71}, 134507 (2005).

%
\bibitem{chuanwei-2011}
M. Gong, S. Tewari,  and C. Zhang, 
arXiv:1105.1796 (2011).

%
\bibitem{iskin-2011}
M. Iskin and A. L. Subasi, 
arXiv:1106.0473 (2011).

%
\bibitem{nakahara-1990}
M. Nakahara, 
Adam Hilger, Bristol (1990).

\end{thebibliography}
\end{document}